\documentclass[12pt]{article}
\usepackage{epsfig}
\newcommand\bea{\begin{eqnarray}}
\newcommand\eea{\end{eqnarray}}

\begin{document}
\thispagestyle{empty}
\bibliographystyle{unsrt}
\setlength{\baselineskip}{18pt}
\parindent 24pt
%\vspace{60pt}

\begin{center}{
{\Large \bf Decoherence and asymptotic entanglement \\
in open
quantum dynamics} \vskip 0.5truecm
Aurelian Isar \\
{\it Department of Theoretical Physics \\
National Institute of Physics and Nuclear Engineering\\
Bucharest-Magurele, Romania} \\
e-mail: isar@theory.nipne.ro
}
\end{center}

\begin{abstract}
In the framework of the theory of open systems based on completely positive quantum
dynamical semigroups, we
determine the degree of quantum decoherence of a harmonic oscillator
interacting with a thermal bath. It is found that the system
manifests a quantum decoherence which is more and more significant
in time. We also calculate the decoherence time and show that it has
the same scale as the time after which thermal fluctuations become
comparable with quantum fluctuations. Then we solve the master equation for two independent 
harmonic oscillators interacting with an environment
in the asymptotic long-time regime. We give a description of
the continuous-variable asymptotic entanglement in terms of the covariance
matrix of the quantum states of the considered system for an
arbitrary Gaussian input state. Using the Peres--Simon
necessary and sufficient condition for separability of two-mode
Gaussian states, we show that the two non-interacting systems
immersed in a common environment become asymptotically
entangled for certain environments, so that in the long-time regime they manifest non-local 
quantum correlations.
\end{abstract}


\section{Introduction}

By quantum decoherence (QD) \cite{joo,zur} we understand the
irreversible, uncontrollable and persistent formation of quantum
correlations (entanglement) of a system with its environment
\cite{ali}, expressed by the damping of the coherences present in the
quantum state of the system, when the off-diagonal elements of the
density matrix decay below a certain level, so that this density
matrix becomes approximately diagonal.

Since QD strongly depends on the interaction between the system and its external environment 
\cite{joo,zur}, its role became relevant in many interesting physical problems. In many cases 
one is interested in understanding QD
because one wants to prevent decoherence from damaging quantum
states and to protect the information stored in quantum states from
the degrading effect of the interaction with the environment.
QD is also responsible for washing out the quantum
interference effects which are desirable to be seen as signals in
experiments and it has a negative influence on many areas relying upon
quantum coherence effects, in particular QD is a major problem in
quantum optics and physics of quantum information and computation
\cite{nie}.

Quantum entanglement represents the physical resource in
quantum information science which is indispensable for the
description and performance of such tasks like teleportation,
superdense coding, quantum cryptography and quantum computation
\cite{nie}. Therefore the generation, detection and
manipulation of the entanglement continues to be presently a
problem of intense investigation.

When two systems are immersed in an environment, then, besides and at the same time with the
QD phenomenon, the external environment can also generate a quantum entanglement
of the two systems and therefore an additional mechanism to
correlate them \cite{ben2,ben3,dod3}. In certain circumstances,
the environment enhances entanglement and in others it
suppresses the entanglement and the state describing the two
systems becomes separable. The structure
and properties of the environment may be such that not only the
two systems become entangled, but also such that a certain
amount of entanglement survives in the asymptotic long-time
regime. The reason is that even if not directly coupled, the
two systems immersed in the same environment can interact
through the environment itself and it depends on how strong
this indirect interaction is with respect to the QD,
whether entanglement can be generated at the beginning of the
evolution and, in the case of an affirmative answer, if it can
be maintained for a definite time or it survives indefinitely
in time \cite{ben2}.

In this work we study QD of a harmonic oscillator interacting with
an environment, in particular with a thermal bath, in the framework
of the theory of open quantum systems based on completely positive dynamical semigroups. 
We determine the
degree of QD \cite{mor} for different regimes of the temperature of environment. It is found 
that the system manifests a QD which in general increases with time and temperature. 
We also calculate the decoherence time and show that it has
the same scale as the time after which thermal fluctuations become
comparable with quantum fluctuations.Then we investigate, in the same framework, 
the existence of the continuous variable asymptotic entanglement for a subsystem composed of
two identical harmonic oscillators interacting with an
environment.  We
are interested in discussing the correlation effect of the
environment, therefore we assume that the two systems are
independent, i.e. they do not interact directly. The initial
state of the subsystem is taken of Gaussian form and the
evolution under the quantum dynamical semigroup assures the
preservation in time of the Gaussian form of the state.
We only
investigate here the asymptotic behaviour of the subsystem
states. The time evolution of the entanglement, in particular
the possibility of the so-called "entanglement sudden death",
that is suppression of the entanglement at a certain finite
moment of time, will be discussed in a future work.

The organizing of the paper is as follows. In Sec. 2 we review
the Markovian master equation for the damped harmonic oscillator
and solve it in coordinate representation.
Then in Sec. 3 we investigate QD and calculate the
decoherence time of the system.
In Sec. 4 we write the equations of motion in the Heisenberg
picture for two independent harmonic oscillators interacting with a general environment. 
With these equations we
derive in Sec. 5 the asymptotic values of the variances and
covariances of the coordinates and momenta which enter the
asymptotic covariance matrix. Then, by using the
Peres-Simon necessary and sufficient condition for separability
of two-mode Gaussian states \cite{per,sim}, we investigate the
behaviour of the environment induced entanglement in the limit
of long times. We show that for certain classes of environments
the initial state evolves asymptotically to an equilibrium
state which is entangled, while for other values of the
parameters describing the environment, the entanglement is
suppressed and the asymptotic state is separable. The existence
of the quantum correlations between the two systems in the
asymptotic long-time regime is the result of the competition
between entanglement and QD. A summary is given in
Sec. 6.

\section{Markovian master equation for a harmonic oscillator}

In the axiomatic formalism based on completely positive quantum dynamical
semigroups, the irreversible time evolution of an open system is
described by the following general quantum Markovian master equation
for the density operator $\rho(t)$ \cite{lin}:
\begin{eqnarray}{d \rho(t)\over dt}=-{i\over\hbar}[H,\rho(t)]
+{1\over 2\hbar} \sum_{j}([  V_{j} \rho(t), V_{j}^\dagger ]+[ V_{j},
\rho(t) V_{j}^\dagger ]).\label{lineq}\end{eqnarray} The harmonic
oscillator Hamiltonian $H$ is chosen of the general quadratic form
\begin{eqnarray} H=H_{0}+{\mu\over 2}(xp_x+p_xx), ~~~  H_{0}={1\over
2m}p_x^2+{m\omega^2\over 2}x^2 \label{ham} \end{eqnarray} and the
operators $V_{j},$ $ V_{j}^\dagger,$ which model the environment,
are taken as linear polynomials in coordinate $x$ and momentum $p_x.$
Then the master equation (\ref{lineq}) takes the following form
\cite{rev}:
\begin{eqnarray} {d \rho \over dt}=-{i\over \hbar}[ H_{0}, \rho]-
{i\over 2\hbar}(\lambda +\mu) [x, \rho p_x+ p_x \rho]+{i\over
2\hbar}(\lambda -\mu)[  p_x,
\rho  x+  x \rho]  \nonumber\\
-{D_{p_xp_x}\over {\hbar}^2}[x,[x, \rho]]-{D_{xx}\over {\hbar}^2} [
p_x,[  p_x, \rho]]+{D_{xp_x}\over {\hbar}^2}([x,[  p_x, \rho]]+ [ p_x,[x,
\rho]]). ~~~~\label{mast}   \end{eqnarray} The diffusion
coefficients $D_{xx},D_{p_xp_x},$ $D_{xp_x}$ and the dissipation constant
$\lambda$ satisfy the fundamental constraints: $ D_{xx}>0, D_{p_xp_x}>0$
and \bea D_{xx}D_{p_xp_x}-D_{xp_x}^2\ge \frac{{\lambda}^2{\hbar}^2}{4}.\eea In the
particular case when the asymptotic state is a Gibbs state $
\rho_G(\infty)=e^{-{  H_0\over kT}}/ {\rm Tr}e^{-{ H_0\over kT}}, $
these coefficients become
\begin{eqnarray} D_{xx}={\lambda-\mu\over
2}{\hbar\over m\omega}\coth{\hbar\omega\over 2kT},~~D_{p_xp_x}={\lambda+\mu\over 2}\hbar
m\omega\coth{\hbar\omega\over 2kT},~~D_{xp_x}=0,
\label{coegib}
\end{eqnarray} where $T$ is the temperature of the thermal bath. In
this case, the fundamental constraints are satisfied only if
$\lambda>\mu$ and
\begin{eqnarray} (\lambda^2-\mu^2)\coth^2{\hbar\omega\over 2kT}
\ge\lambda^2\label{cons}\end{eqnarray} and the asymptotic values
$\sigma_{xx}(\infty),$ $\sigma_{p_xp_x}(\infty),$ $\sigma_{xp_x}(\infty)$
of the dispersion (variance), respectively correlation (covariance),
of the coordinate and momentum, reduce to \cite{rev}
\begin{eqnarray} \sigma_{xx}(\infty)={\hbar\over
2m\omega}\coth{\hbar\omega\over 2kT}, ~~\sigma_{p_xp_x}(\infty)={\hbar
m\omega\over 2}\coth{\hbar\omega\over 2kT}, ~~\sigma_{xp_x}(\infty)=0.
\label{varinf} \end{eqnarray}

We consider a harmonic oscillator with an initial Gaussian wave
function ($\sigma_x(0)$ and $\sigma_{p_x}(0)$ are the initial averaged
position and momentum of the wave packet) \begin{eqnarray}
\Psi(x)=({1\over 2\pi\sigma_{xx}(0)})^{1\over 4}\exp[-{1\over
4\sigma_{xx}(0)}
(1-{2i\over\hbar}\sigma_{xp_x}(0))(x-\sigma_x(0))^2+{i\over
\hbar}\sigma_{p_x}(0)x], \label{ccs}\end{eqnarray} representing a
correlated coherent state \cite{dod1} (squeezed coherent state) with the
variances and covariance of coordinate and momentum
\begin{eqnarray} \sigma_{xx}(0)={\hbar\delta\over 2m\omega},~~
\sigma_{p_xp_x}(0)={\hbar m\omega\over 2\delta(1-r^2)},~~
\sigma_{xp_x}(0)={\hbar r\over 2\sqrt{1-r^2}}.
\label{inw}\end{eqnarray} Here $\delta$ is the squeezing parameter
which measures the spread in the initial Gaussian packet and $r,$
with $|r|<1$ is the correlation coefficient. The initial values
(\ref{inw}) correspond to a minimum uncertainty state, since they
fulfil the generalized uncertainty relation
\bea\sigma_{xx}(0)\sigma_{p_xp_x}(0)-\sigma_{xp_x}^2(0) =\frac{\hbar^2}{4}.\eea For
$\delta=1$ and $r=0$ the correlated coherent state becomes a Glauber
coherent state.

From Eq. (\ref{mast}) we derive the evolution equation in coordinate
representation: \begin{eqnarray} {\partial\rho\over\partial
t}={i\hbar\over 2m}({\partial^2\over\partial x^2}-
{\partial^2\over\partial x'^2})\rho-{im\omega^2\over
2\hbar}(x^2-x'^2)\rho\nonumber\\
-{1\over 2}(\lambda+\mu)(x-x')({\partial\over\partial
x}-{\partial\over\partial x'})\rho+{1\over
2}(\lambda-\mu)[(x+x')({\partial\over\partial
x}+{\partial\over\partial
x'})+2]\rho  \nonumber\\
-{D_{p_xp_x}\over\hbar^2}(x-x')^2\rho+D_{xx}({\partial\over\partial
x}+{\partial\over \partial x'})^2\rho -{2iD_{xp_x}\hbar}(x-x')(
{\partial\over\partial x}+{\partial\over\partial
x'})\rho.\label{cooreq}\end{eqnarray} The first two terms on the
right-hand side of this equation generate the usual Liouvillian
unitary evolution. The third and forth terms are the dissipative
terms and have a damping effect (exchange of energy with
environment). The last three are noise (diffusive) terms and produce
fluctuation effects in the evolution of the system. $D_{p_xp_x}$
promotes diffusion in momentum and generates decoherence in
coordinate $x$ -- it reduces the off-diagonal terms, responsible for
correlations between spatially separated pieces of the wave packet.
Similarly $D_{xx}$ promotes diffusion in coordinate and generates
decoherence in momentum $p_x.$ The $D_{xp_x}$ term is the so-called
"anomalous diffusion" term and it does not generate decoherence.

The density matrix solution of Eq. (\ref{cooreq}) has the general
Gaussian form \begin{eqnarray} <x|\rho(t)|x'>=({1\over
2\pi\sigma_{xx}(t)})^{1\over 2} \exp[-{1\over
2\sigma_{xx}(t)}({x+x'\over
2}-\sigma_x(t))^2\nonumber\\
-{\sigma(t)\over 2\hbar^2\sigma_{xx}(t)}(x-x')^2
+{i\sigma_{xp_x}(t)\over \hbar\sigma_{xx}(t)}({x+x'\over
2}-\sigma_x(t))(x-x')+{i\over
\hbar}\sigma_{p_x}(t)(x-x')],\label{densol} \end{eqnarray} where
\bea\sigma(t)\equiv\sigma_{xx}(t)\sigma_{p_xp_x}(t)-\sigma_{xp_x}^2(t)\eea is
the determinant of the covariance matrix
\bea \pmatrix{ \sigma_{xx}(t) & \sigma_{xp_x}(t) \cr
\sigma_{xp_x}(t) & \sigma_{p_xp_x}(t)}\eea and represents also the
Schr\"odinger generalized uncertainty function. In the case of a
thermal bath we obtain the following stationary state solution for
$t\to\infty$ ($\epsilon\equiv{\hbar\omega/2kT}$):
\begin{eqnarray} <x|\rho(\infty)|x'>=({m\omega\over
\pi\hbar\coth\epsilon})^{1\over 2}\exp\{-{m\omega\over
4\hbar}[{(x+x')^2\over\coth\epsilon}+
(x-x')^2\coth\epsilon]\}.\label{dinf}\end{eqnarray}

\section{Quantum decoherence}

An isolated system has an unitary evolution and the coherence of the
state is not lost -- pure states evolve in time only to pure states.
The QD phenomenon, that is the loss of coherence or the destruction
of off-diagonal elements representing coherences between quantum
states in the density matrix, can be achieved by introducing an
interaction between the system and environment: an initial pure
state with a density matrix which contains nonzero off-diagonal
terms can non-unitarily evolve into a final mixed state with a
diagonal density matrix.

In the literature several measures of the degree of decoherence have been introduced, 
expressed in terms of the von Neumann entropy \cite{miz}, linear entropy \cite{zur1,a4} 
or the coefficients of the matrix elements of the statistical operator with respect 
to the energy basis \cite{dod2}.

Using new variables $\Sigma=(x+x')/2$ and $\Delta=x-x',$ the density
matrix (\ref{densol}) becomes \begin{eqnarray}
\rho(\Sigma,\Delta,t)=\sqrt{\alpha\over \pi}\exp[-\alpha\Sigma^2
-\gamma\Delta^2
+i\beta\Sigma\Delta+2\alpha\sigma_x(t)\Sigma+i({\sigma_{p_x}(t)\over\hbar}-
\beta\sigma_x(t))\Delta-\alpha\sigma_x^2(t)],\label{ccd3}\end{eqnarray}
with the abbreviations \begin{eqnarray} \alpha={1\over
2\sigma_{xx}(t)},~~\gamma={\sigma(t)\over 2\hbar^2
\sigma_{xx}(t)},~~ \beta={\sigma_{xp_x}(t)\over\hbar\sigma_{xx}(t)}.
\label{ccd4}\end{eqnarray}

In the present work we use the representation-independent measure of the degree of QD \cite{mor}, which
is given by the ratio of the dispersion $1/\sqrt{2\gamma}$ of the
off-diagonal element $\rho(0,\Delta,t)$ to the dispersion
$\sqrt{2/\alpha}$ of the diagonal element $\rho(\Sigma,0,t):$
\begin{eqnarray} \delta_{QD}(t)={1\over 2}\sqrt{\alpha\over
\gamma}={\hbar\over 2\sqrt{\sigma(t)}}.\label{qdec}\end{eqnarray}
It can easily be shown that $\delta_{QD}$ is related to the linear entropy \cite{a3}.

The finite temperature Schr\"odinger generalized uncertainty
function has the expression \cite{a1} (with the notation
$\Omega^2\equiv\omega^2-\mu^2$,
$\omega>\mu$)\begin{eqnarray}\sigma(t)={\hbar^2\over
4}\{e^{-4\lambda
t}[1-(\delta+{1\over\delta(1-r^2)})\coth\epsilon+\coth^2\epsilon]\nonumber\\
+e^{-2\lambda t}\coth\epsilon[(\delta+{1\over\delta(1-r^2)}
-2\coth\epsilon){\omega^2-\mu^2\cos(2\Omega
t)\over\Omega^2}\nonumber\\ +(\delta-{1\over\delta(1-r^2)}){\mu
\sin(2\Omega t)\over\Omega}+{2r\mu\omega(1-\cos(2\Omega
t))\over\Omega^2\sqrt{1-r^2}}]+\coth^2\epsilon\}.\label{sunc}\end{eqnarray}
In the limit of long times Eq. (\ref{sunc}) yields
\bea\sigma(\infty)=\frac{\hbar^2}{4}\coth^2\epsilon,\eea so that we obtain
\begin{eqnarray} \delta_{QD}(\infty)=\tanh{\hbar\omega\over
2kT},\end{eqnarray} which for high $T$ becomes
\bea\delta_{QD}(\infty)=\frac{\hbar\omega}{2kT}.\eea To illustrate the dependence
on temperature and time of the degree of QD, we represent it in Fig.
1. We see that in general $\delta_{QD}$ decreases, and therefore QD
becomes stronger, with time and temperature, i.e. the density matrix
becomes more and more diagonal at higher $T$ and the contributions
of the off-diagonal elements get smaller and smaller. At the same
time the degree of purity decreases and the degree of mixedness
increases with $T.$ For $T=0$ the asymptotic (final) state is pure
and $\delta_{QD}$ reaches its initial maximum value 1. $\delta_{QD}=
0$ when the quantum coherence is completely lost, and if
$\delta_{QD}= 1$ there is no QD. Only if $\delta_{QD}<1$ we can say
that the considered system interacting with the thermal bath
manifests QD, when the magnitude of the elements of the density
matrix in the position basis are peaked preferentially along the
diagonal $x=x'.$ In Fig. 2 we represent the density matrix in
coordinate representation (\ref{densol}) at the initial and final
moments of time. The values of density matrix along the diagonal
$x=x'$ represent the probability of finding the system in this
position, while off-diagonal values represent the correlations in
the density matrix between the points $x$ and $x'.$ For simplicity,
in Fig. 2 we consider zero values for the initial expectations
values of coordinate and momentum, so that the density matrix is
centered in origin. Dissipation promotes quantum coherences, whereas
fluctuation (diffusion) reduces coherences and promotes QD. The
balance of dissipation and fluctuation determines the final
equilibrium value of $\delta_{QD}$ \cite{a2}.

\begin{figure}
\label{Fig1}
\centerline{\epsfig{file=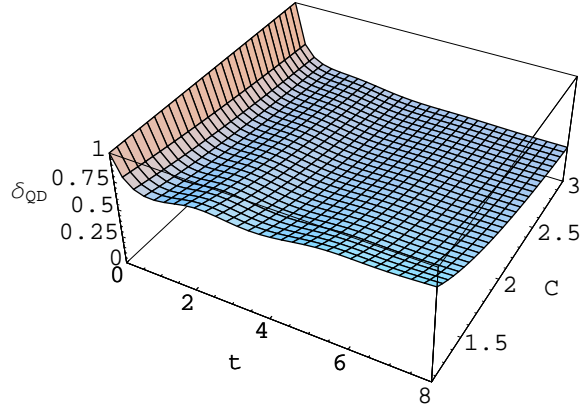,width=0.5\textwidth}}
\caption{Degree of quantum decoherence $\delta_{QD}$ versus
temperature $T$ (through $C\equiv\coth{(\hbar\omega/2kT)}$) and time
$t$ for $\lambda=0.2,\mu=0.1,\delta=4,r=0.$}
\end{figure}

\begin{figure}
\label{Fig2}
\centerline{\epsfig{file=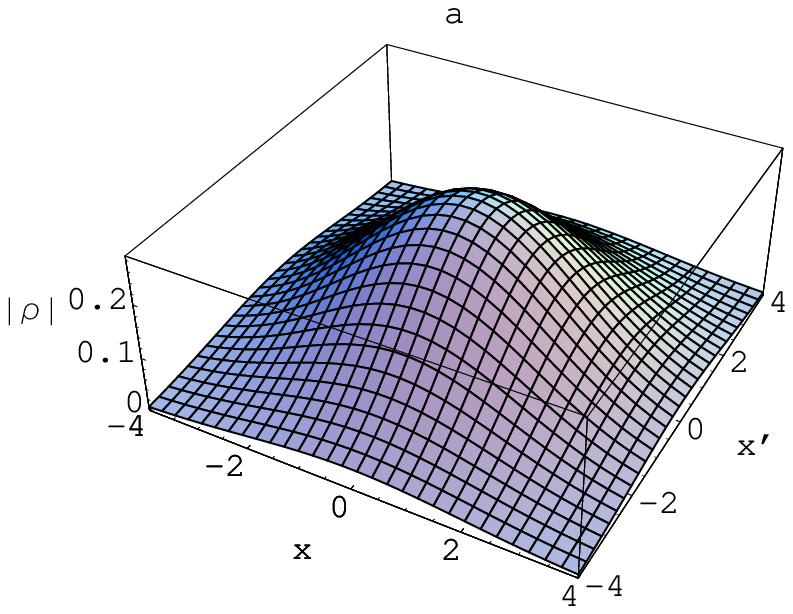,width=0.5\textwidth}
\epsfig{file=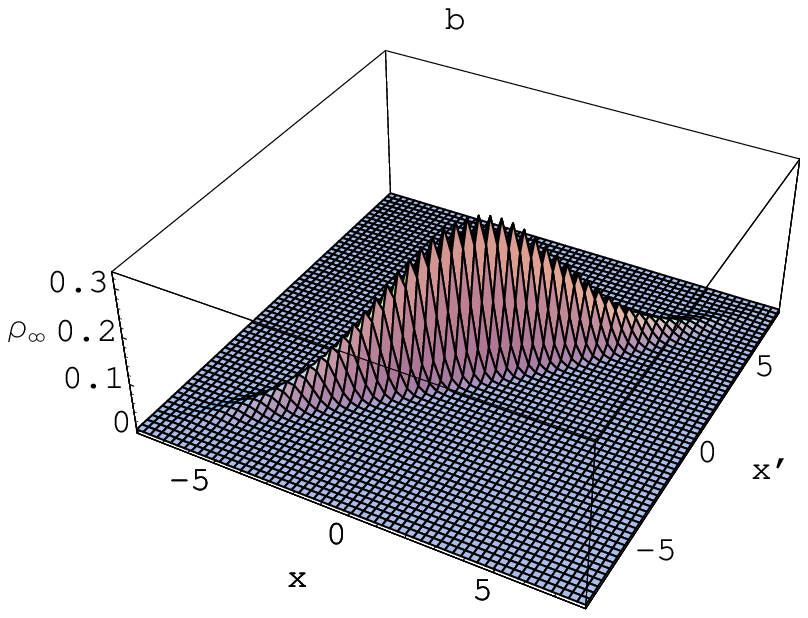,width=0.5\textwidth}}
%,keepaspectratio=true]
\caption{Density matrix $\rho$ in coordinate representation for
$\lambda=0.2,\mu=0.1,\delta=4,r=0:$ (a) $|\rho|$ at the initial time
$t=0;$ (b) $\rho_{\infty}$ for $C=10.$}
\end{figure}

In order to obtain the expression of the decoherence time, we
consider the coefficient $\gamma$ (\ref{ccd4}), which measures the
contribution of non-diagonal elements in the density matrix
(\ref{ccd3}). For short times ($\lambda t\ll 1, \Omega t\ll 1$), we
have: \bea \gamma(t)=-{m\omega\over
4\hbar\delta}\{1+2[\lambda(\delta+{r^2\over\delta(1-r^2)})\coth\epsilon
+\mu(\delta-{r^2\over\delta(1-r^2)})\coth\epsilon-\lambda-\mu-{\omega
r\over\delta\sqrt{1-r^2}}]t\}.\label{td}\eea From here we deduce that the quantum coherences
in the density matrix decay exponentially and the decoherence time
scale is given by \bea t_{deco}=\{
2[\lambda(\delta+\displaystyle{{r^2\over\delta(1-r^2)}})\coth\epsilon
+\mu(\delta-\displaystyle{{r^2\over\delta(1-r^2)}})\coth\epsilon
-\lambda-\mu-\displaystyle{{\omega
r\over\delta\sqrt{1-r^2}}}]\}^{-1}.\label{tdeco1}\eea The decoherence time
depends on the temperature $T$ and the coupling $\lambda$
(dissipation coefficient) between the system and environment, the
squeezing parameter $\delta$ and the initial correlation coefficient
$r.$ We notice that the decoherence time is decreasing with
increasing dissipation, temperature and squeezing.

For $r=0$ we obtain:\bea t_{deco}={1\over
2(\lambda+\mu)(\delta\coth\epsilon-1)}\label{tdeco2}\eea and at
temperature $T=0$ [here we have to take $\mu=0$ due to Eq. (\ref{cons})], this becomes \bea
t_{deco}={1\over 2\lambda(\delta-1)}.\eea We see that when the
initial state is the usual coherent state $(\delta=1),$ then the
decoherence time tends to infinity. This corresponds to the fact
that for $T=0$ and $\delta=1$ the coefficient $\gamma$ is constant
in time, so that the decoherence process does not occur in this
case.

At high temperature, expression (\ref{tdeco1}) becomes \bea t_{deco}=\{
2[\lambda(\delta+\displaystyle{{r^2\over\delta(1-r^2)}})
+\mu(\delta-{r^2\over\delta(1-r^2)})]\frac
{2kT}{\hbar\omega}\}^{-1}.\eea If, in addition
$r=0,$ then we obtain \bea t_{deco}={\hbar\omega\over
4(\lambda+\mu)\delta kT}.\eea

The generalized uncertainty function $\sigma(t)$ (\ref{sunc}) has
the following behaviour for short times: \bea
\sigma(t)={\hbar^2\over 4}\{1+ 2[\lambda
(\delta+{1\over\delta(1-r^2)})\coth\epsilon+\mu(\delta-{1\over\delta(1-r^2)})
\coth\epsilon-2\lambda]t\}.\label{sunc1}\eea This expression shows
explicitly the contribution for small time of uncertainty that is
intrinsic to quantum mechanics, expressed through the Heisenberg
uncertainty principle and uncertainty due to the coupling to the
thermal environment. From Eq. (\ref{sunc1}) we can determine the
time $t_d$ when thermal fluctuations become comparable with quantum
fluctuations. At high temperature we obtain \bea t_d=\{
2[\lambda
(\delta+\displaystyle{{1\over\delta(1-r^2)}})+\mu(\delta-\displaystyle{
{1\over\delta(1-r^2)}})]\frac
{2kT}{\hbar\omega}\}^{-1}.\eea
As expected, the decoherence time $t_{deco}$ has the
same scale as the time $t_d$ after which thermal fluctuations become
comparable with quantum fluctuations \cite{a2,hu}. The values of
$t_{deco}$ and $t_d$ become closer with increasing temperature and
squeezing.

When $t\gg t_{rel},$ where $t_{rel}\approx\lambda^{-1}$ is the
relaxation time, which governs the rate of energy dissipation, the
particle reaches equilibrium with the environment. In the
macroscopic domain QD occurs very much faster than relaxation, so
that for all macroscopic bodies the dissipation term becomes
important much later after the decoherence term has already
dominated and diminished the off-diagonal terms of the density
matrix. We remark also that $t_{deco}$ can be of the order of
$t_{rel}$ for sufficiently low temperatures and small wave packet
spread (small squeezing coefficient).

\section{Equations of motion for two independent harmonic oscillators}

We are now interested in the generation of entanglement between two
harmonic oscillators due to the back-action of the environment
on the subsystem. Since the two harmonic oscillators interact
with a common environment, there will be induced coupling
between the two oscillators even when initially they are
uncoupled. Thus, the master equation for the two harmonic oscillators must account 
for their mutual interaction by their
coupling to the environment. We shall study the dynamics
of the subsystem composed of two identical non-interacting
(independent) oscillators in weak interaction with a large
environment, so that their reduced time evolution can be
described by a Markovian, completely positive quantum dynamical
semigroup, like in the previous case of one harmonic oscillator.

If $ \widetilde \Phi_t $ is the dynamical semigroup describing
the time evolution of the open quantum system in the Heisenberg
picture ($ \widetilde \Phi_t $ is the dual of the dynamical semigroup $ \Phi_t $ which 
describes the time evolution in the Schr\"odinger picture given by Eq. (\ref{lineq})), 
then the master equation is given for an operator $A$
as follows \cite{lin,rev}: \bea{d\widetilde\Phi_t(A)\over
dt}={i\over \hbar}[H,\widetilde\Phi_t(A)]+{1\over
2\hbar}\sum_j(V_j^{\dagger}[\widetilde\Phi_t(A),
V_j]+[V_j^{\dagger},\widetilde\Phi_t(A)]V_j).\label{masteq}\eea
Here, again $H$ denotes the Hamiltonian of the open quantum system
and $V_j, V_j^\dagger$ which are operators defined on the Hilbert space of $H,$
represent the interaction of the open system
with the environment. Being interested
in the set of Gaussian states, we introduce those quantum
dynamical semigroups that preserve that set. Therefore $H$ is
taken to be a polynomial of second degree in the coordinates
$x,y$ and momenta $p_x,p_y$ of the two quantum oscillators and
$V_j,V_j^{\dagger}$ are taken polynomials of only first degree
in these canonical observables. Then in the linear space
spanned by the coordinates and momenta there exist only four
linearly independent operators $V_{j=1,2,3,4}$ \cite{san}: \bea
V_j=a_{xj}p_x+a_{yj}p_y+b_{xj}x+b_{yj}y,\eea where
$a_{xj},a_{yj},b_{xj},b_{yj}\in {\bf C}$ and \bea
V_j^{\dagger}=a_{xj}^*p_x+a_{yj}^*p_y+b_{xj}^*x+b_{yj}^*y,\eea
where $*$ denotes the complex conjugation.
The Hamiltonian $H$ of the two uncoupled identical harmonic
oscillators of mass $m$ and frequency $\omega$ is given by \bea H={1\over 2m}(p_x^2+p_y^2)+{m\omega^2\over
2}(x^2+y^2).\eea

The fact that $\widetilde \Phi_t$ is a dynamical semigroup
implies the positivity of the following matrix formed by the
scalar products of the four vectors $ {\bf a}_x, {\bf a}_y,
{\bf b}_x, {\bf b}_y$ whose entries are the components $a_{xj},a_{yj},b_{xj},b_{yj},$ respectively:
\begin{eqnarray}{1 \over 2} \hbar \pmatrix{
({\bf a}_x {\bf a}_x)&({\bf a}_x {\bf b}_x) &({\bf a}_x {\bf
a}_y)&({\bf a}_x {\bf b}_y) \cr ({\bf b}_x {\bf a}_x)&({\bf
b}_x {\bf b}_x) &({\bf b}_x {\bf a}_y)&({\bf b}_x {\bf b}_y)
\cr ({\bf a}_y {\bf a}_x)&({\bf a}_y {\bf b}_x) &({\bf a}_y
{\bf a}_y)&({\bf a}_y {\bf b}_y) \cr ({\bf b}_y {\bf
a}_x)&({\bf b}_y {\bf b}_x) &({\bf b}_y {\bf a}_y)&({\bf b}_y
{\bf b}_y)}.
\end{eqnarray}
For simplicity we take this matrix of the following form, where
all coefficients $D_{xx}, D_{xp_x},$... and $\lambda$ are real quantities:
\bea \pmatrix{ D_{xx}&- D_{xp_x} - i \hbar \lambda/2&D_{xy}& -
D_{xp_y} \cr - D_{xp_x} + i \hbar\lambda/2&D_{p_x p_x}&-
D_{yp_x}&D_{p_x p_y} \cr D_{xy}&- D_{y p_x}&D_{yy}&- D_{y p_y}
- i \hbar \lambda/2 \cr - D_{xp_y} &D_{p_x p_y}&- D_{yp_y} + i
\hbar \lambda/2&D_{p_y p_y} }.\label{coef} \eea It follows that
the principal minors of this matrix are positive or zero. From
the Cauchy-Schwarz inequality the following relations for the
coefficients defined in Eq. (\ref{coef}) hold (from now on we
put, for simplicity, $\hbar=1$): \bea D_{xx}D_{yy}-D^2_{xy}\ge
0,~ D_{xx}D_{p_xp_x}-D^2_{xp_x}\ge\frac{\lambda^2}{4},~
D_{xx}D_{p_yp_y}-D^2_{xp_y}\ge 0,~\\
D_{yy}D_{p_xp_x}-D^2_{yp_x}\ge 0,~
D_{yy}D_{p_yp_y}-D^2_{yp_y}\ge\frac{\lambda^2}{4},~
D_{p_xp_x}D_{p_yp_y}-D^2_{p_xp_y}\ge 0.\eea

The matrix of the coefficients (\ref{coef}) can be conveniently
written as \bea \pmatrix{ {C_1} & {C_3} \cr
 {C_3}^{\dagger} & {C_2}},
\label{subm} \eea in terms of $2\times 2$ matrices
$C_1={C_1}^\dagger$, $C_2={C_2}^\dagger$ and ${C_3}$. This
decomposition has a direct physical interpretation: the
elements containing the diagonal contributions $C_1$ and $C_2$
represent diffusion and dissipation coefficients corresponding
to the first, respectively the second, system in absence of the
other, while the elements in $C_3$ represent environment
generated couplings between the two, initially independent,
oscillators.

The variance and covariance of self-adjoint operators $A_1$ and
$A_2$ can be written with the density operator $\rho$,
describing the initial state of the quantum system, as follows:
\bea\sigma_{A_1A_2}(t)={1\over 2}{\rm
Tr}(\rho\widetilde\Phi_t(A_1A_2+A_2A_1)).\eea

We introduce the following $4\times 4$ covariance matrix:
\bea\sigma(t)=\pmatrix{\sigma_{xx}&\sigma_{xp_x} &\sigma_{xy}&
\sigma_{xp_y}\cr \sigma_{xp_x}&\sigma_{p_xp_x}&\sigma_{yp_x}
&\sigma_{p_xp_y}\cr \sigma_{xy}&\sigma_{yp_x}&\sigma_{yy}
&\sigma_{yp_y}\cr \sigma_{xp_y}&\sigma_{p_xp_y}&\sigma_{yp_y}
&\sigma_{p_yp_y}}.\label{covar} \eea

By direct calculation we obtain \cite{san}: \bea{d \sigma\over
dt} = Y \sigma + \sigma Y^{\rm T}+2 D,\label{vareq}\eea where
\bea Y=\pmatrix{ -\lambda&1/m&0 &0\cr -m\omega^2&-\lambda&0&
0\cr 0&0&-\lambda&1/m \cr 0&0&-m\omega^2&-\lambda}, \eea $D$ is
the matrix of the diffusion coefficients \bea D=\pmatrix{
D_{xx}& D_{xp_x} &D_{xy}& D_{xp_y} \cr D_{xp_x}&D_{p_x p_x}&
D_{yp_x}&D_{p_x p_y} \cr D_{xy}& D_{y p_x}&D_{yy}& D_{y p_y}
\cr D_{xp_y} &D_{p_x p_y}& D_{yp_y} &D_{p_y p_y} }\eea and
$Y^{\rm T}$ is the transposed matrix of $Y$. The time-dependent
solution of Eq. (\ref{vareq}) is given by \cite{san}
\bea\sigma(t)= M(t)(\sigma(0)-\sigma(\infty)) M^{\rm
T}(t)+\sigma(\infty),\eea where $M(t)=\exp(tY).$ The matrix
$M(t)$ has to fulfil the condition $\lim_{t\to\infty} M(t)=0.$
In order that this limit exists, $Y$ must only have eigenvalues
with negative real parts. The values at infinity are obtained
from the equation \cite{san} \bea
Y\sigma(\infty)+\sigma(\infty) Y^{\rm T}=-2 D.\label{covarinf}\eea

\section{Environment induced entanglement}

The two-mode Gaussian state is entirely specified by its
covariance matrix $\sigma$ (\ref{covar}), which is a real,
symmetric and positive matrix with the following block
structure:
\begin{eqnarray}
\sigma=\left(\begin{array}{cc}A&C\\
C^{\rm T}&B \end{array}\right),
\end{eqnarray}
where $A$, $B$ and $C$ are $2\times 2$ matrices. Their entries
are correlations of the canonical operators $x,y,p_x,p_y$, $A$
and $B$ denote the symmetric covariance matrices for the
individual reduced one-mode states, while the matrix $C$
contains the cross-correlations between modes. The entries of
the covariance matrix depend on $Y$ and $D$ and can be
calculated from Eq. (\ref{covarinf}). To simplify further the
calculations, we shall consider environments for which the two
diagonal submatrices in Eq. (\ref{subm}) are equal: $C_1=C_2$,
so that $D_{xx}=D_{yy}, D_{xp_x}=D_{yp_y},
D_{p_xp_x}=D_{p_yp_y}.$ In addition, in the matrix $C_3$ we
take $D_{xp_y}=D_{yp_x}.$ Then both unimodal covariance
matrices are equal, $A=B$ and the entanglement matrix $C$ is
symmetric. With the chosen coefficients, we obtain the
following elements of the asymptotic entanglement matrix $C$:
\bea\sigma_{xy} (\infty) =
\frac{m^2(2\lambda^2+\omega^2)D_{xy}+2m\lambda
D_{xp_y}+D_{p_xp_y}} {2m^2\lambda(\lambda^2+\omega^2)},\eea
\bea\sigma_{xp_y}(\infty)=
\sigma_{yp_x}(\infty)=\frac{-m^2\omega^2 D_{xy}+2m\lambda
D_{xp_y}+ D_{p_xp_y}}{2m(\lambda^2+\omega^2)},\eea
\bea\sigma_{p_xp_y} (\infty) =
\frac{m^2\omega^4D_{xy}-2m\omega^2\lambda D_{xp_y}+(2\lambda^2
+\omega^2)D_{p_xp_y}}{2\lambda(\lambda^2+\omega^2)}\eea and of the matrices $A$ and $B$: 
\bea\sigma_{xx}
(\infty) =\sigma_{yy} (\infty)=
\frac{m^2(2\lambda^2+\omega^2)D_{xx}+ 2m\lambda
D_{xp_x}+D_{p_xp_x}}{2m^2\lambda(\lambda^2+\omega^2)},\eea
\bea\sigma_{xp_x}(\infty)=
\sigma_{yp_y}(\infty)=\frac{-m^2\omega^2 D_{xx}+2m\lambda
D_{xp_x}+ D_{p_xp_x}}{2m(\lambda^2+\omega^2)},\eea
\bea\sigma_{p_xp_x} (\infty) = \sigma_{p_yp_y}
(\infty)=\frac{m^2\omega^4D_{xx}-2m\omega^2 \lambda
D_{xp_x}+(2\lambda^2+\omega^2)D_{p_xp_x}}
{2\lambda(\lambda^2+\omega^2)}.\eea With these quantities we
calculate the determinant of the entanglement matrix: \bea \det
C=\frac{1}{4\lambda^2(\lambda^2+\omega^2)}[(m\omega^2D_{xy}+
\frac{1}{m}
D_{p_xp_y})^2+4\lambda^2(D_{xy}D_{p_xp_y}-D_{xp_y}^2)].\eea It
is very interesting that the general theory of open quantum
systems allows couplings via the environment between uncoupled
oscillators. According to the definitions of the environment
parameters, the diffusion coefficients above can be different
from zero and can simulate an interaction between the uncoupled
oscillators. Indeed, the Gaussian states with $\det C\ge 0$ are
separable states, but for $\det C <0,$ it may be possible that
the asymptotic equilibrium states are entangled, as it will be
shown in the following.

On general grounds, one expects that the effects of
decoherence, counteracting entanglement production, be dominant
in the long-time regime, so that no quantum correlation (entanglement)
is expected to be left at infinity. Nevertheless,
there are situations in which the environment allows the
presence of entangled asymptotic equilibrium states. In order
to investigate whether an external environment can actually
entangle the two independent systems, we can use the partial
transposition criterion \cite{per,sim}: a state results
entangled if and only if the operation of partial transposition
does not preserve its positivity. Simon \cite{sim} obtained the
following necessary and sufficient criterion for separability:
$S\ge 0,$ where \bea S\equiv\det A \det B+(\frac{1}{4} -|\det
C|)^2- {\rm Tr}[AJCJBJC^{\rm T}J]- \frac{1}{4}(\det A+\det B)
\label{sim1}\eea and $J$ is the $2\times 2$ symplectic matrix
\bea
J=\left(\begin{array}{cc}0&1\\
-1&0\end{array}\right).
\end{eqnarray}

In order to analyze the possible persistence of the environment
induced entanglement in the asymptotic long-time regime, we
consider the environment characterized by the following values
of its parameters: $m^2\omega^2D_{xx}=D_{p_xp_x},~~D_{xp_x}=0,
~m^2\omega^2D_{xy}=D_{p_xp_y}.$ In this case the Simon
expression (\ref{sim1}) takes the form: \bea S=
\left(\frac{m^2\omega^2(D_{xx}^2-D_{xy}^2)}{\lambda^2}+
\frac{D_{xp_y}^2}{\lambda^2+\omega^2}-\frac{1}{4}\right)^2-4\frac
{m^2\omega^2D_{xx}^2D_{xp_y}^2}{\lambda^2(\lambda^2+
\omega^2)}.\label{sim2}\eea For environments characterized by
such coefficients that the expression (\ref{sim2}) is negative,
the asymptotic final state is entangled. In particular, if
$D_{xy}=0,$ we obtain that $S<0,$ i.e. the asymptotic final
state is entangled, for the following range of values of the
coefficient $D_{xp_y}$ characterizing the environment: \bea
\frac{m\omega
D_{xx}}{\lambda}-\frac{1}{2}<\frac{D_{xp_y}}{\sqrt{\lambda^2
+\omega^2}}<\frac{m\omega
D_{xx}}{\lambda}+\frac{1}{2},\label{insep}\eea where the
coefficient $D_{xx}$ satisfies the condition $m\omega
D_{xx}/\lambda\ge 1/2,$ equivalent with the unimodal
uncertainty relation. If the coefficients do not fulfil the
inequalities (\ref{insep}), then $S\ge 0$ and therefore the
asymptotic final state of the considered bipartite system is
separable.

\section{Summary}

We have studied QD with the Markovian equation of Lindblad for
a system consisting of an one-dimensional harmonic oscillator
in interaction with a thermal bath in the framework of the
theory of open quantum systems based on completely positive quantum dynamical
semigroups. In the same framework we investigated
the existence of the asymptotic quantum entanglement for a subsystem
composed of two uncoupled identical harmonic oscillators
interacting with an environment.

(1) We have
shown that QD in general increases with time and temperature.
For large temperatures, QD is strong and the degree of
mixedness is high, while for zero temperature the asymptotic
final state is pure. With increasing squeezing parameter and
initial correlation, QD becomes stronger, but the asymptotic
value of the degree of QD does not depend on the initial
squeezing and correlation, it depends on temperature only. QD
is expressed by the loss of quantum coherences in the case of a
thermal bath at finite temperature.

(2) We determined the general expression of the decoherence
time, which shows that it is decreasing with increasing
dissipation, temperature and squeezing. We have also shown that
the decoherence time has the same scale as the time after which
thermal fluctuations become comparable with quantum
fluctuations and the values of these scales become closer with
increasing temperature and squeezing.

(3) By using the Peres-Simon
necessary and sufficient condition for separability of two-mode
Gaussian states, we have shown that for certain classes of
environments the initial state evolves asymptotically to an
equilibrium state which is entangled, i.e. there exist
non-local quantum correlations for the bipartite states of the
two harmonic oscillator subsystem, while for other values of the
coefficients describing the environment, the asymptotic state
is separable.

The obtained results can represent a useful basis for the description of the
connection between uncertainty, decoherence and correlations
(entanglement) of open quantum systems with their environment. Due to the increasing 
interest manifested towards
the continuous variables approach \cite{bra} to the theory of
quantum information, these results, in particular the
possibility of maintaining a bipartite entanglement in a
diffusive-dissipative environment even for asymptotic long
times, could be useful for both phenomenological and
experimental applications in the field of quantum information
processing and communication.

\section*{Acknowledgments}

The author acknowledges the financial support received within
the Project CEEX 68/2005.

\end{document}